\documentclass[sigconf,screen]{acmart}

\usepackage{booktabs}
\usepackage{amsmath}
\usepackage{algorithm}
\usepackage{algorithmic}
\usepackage{graphicx}
\graphicspath{{./}{figures/}}
\usepackage{subcaption}
\usepackage{xcolor}
\usepackage{multirow}

\copyrightyear{2026}
\acmYear{2026}
\setcopyright{cc}
\setcctype{by}
\acmConference[WWW Companion '26]{Companion Proceedings of the ACM Web Conference 2026}{April 13--17, 2026}{Dubai, United Arab Emirates}
\acmBooktitle{Companion Proceedings of the ACM Web Conference 2026 (WWW Companion '26), April 13--17, 2026, Dubai, United Arab Emirates}
\acmPrice{}
\acmDOI{10.1145/3774905.3795728}
\acmISBN{979-8-4007-2308-7/2026/04}

\begin{document}

\title{LLMs as Orchestrators: Constraint-Compliant Multi-Agent Optimization for Recommendation Systems}

\author{Guilin Zhang}
\affiliation{%
  \institution{Workday}
  \city{Pleasanton}
  \state{California}
  \country{USA}
}
\email{guilin.zhang@workday.com}

\author{Kai Zhao}
\affiliation{%
  \institution{Workday}
  \city{Pleasanton}
  \state{California}
  \country{USA}
}
\email{kai.zhao@workday.com}

\author{Jeffrey Friedman}
\affiliation{%
  \institution{Workday}
  \city{Pleasanton}
  \state{California}
  \country{USA}
}
\email{jeffrey.friedman@workday.com}

\author{Xu Chu}
\affiliation{%
  \institution{Workday}
  \city{Pleasanton}
  \state{California}
  \country{USA}
}
\email{xu.chu@workday.com}

\begin{abstract}
Recommendation systems must optimize multiple objectives while satisfying hard business constraints such as fairness and coverage. For example, an e-commerce platform may require every recommendation list to include items from multiple sellers and at least one newly listed product; violating such constraints—even once—is unacceptable in production. Prior work on multi-objective recommendation and recent LLM-based recommender agents largely treat constraints as soft penalties or focus on item scoring and interaction, leading to frequent violations in real-world deployments. How to leverage LLMs for coordinating constrained optimization in recommendation systems remains underexplored. We propose DualAgent-Rec, an LLM-coordinated dual-agent framework for constrained multi-objective e-commerce recommendation. The framework separates optimization into an Exploitation Agent that prioritizes accuracy under hard constraints and an Exploration Agent that promotes diversity through unconstrained Pareto search. An LLM-based coordinator adaptively allocates resources between agents based on optimization progress and constraint satisfaction, while an adaptive $\varepsilon$-relaxation mechanism guarantees feasibility of final solutions. Experiments on the Amazon Reviews 2023 dataset demonstrate that DualAgent-Rec achieves 100\% constraint satisfaction and improves Pareto hypervolume by 4–6\% over strong baselines, while maintaining competitive accuracy–diversity trade-offs. These results indicate that LLMs can act as effective orchestration agents for deployable and constraint-compliant recommendation systems. Our code is publicly available at \url{https://github.com/GuilinDev/Dual-Agents-Recommendation}.
\end{abstract}



\maketitle

\section{Introduction}
\label{sec:intro}

E-commerce recommendation systems play a central role in the web economy and must simultaneously optimize multiple conflicting objectives, including recommendation accuracy, catalog diversity, seller fairness, and new product exposure~\cite{jannach2016recommendations}. To address these competing goals, prior work has explored weighted objective aggregation and multi-objective optimization techniques~\cite{zheng2022multi}. However, in most real-world systems, business requirements such as fairness, coverage, and exposure are enforced only as soft penalties, which often leads to persistent constraint violations and undesirable exposure imbalances in deployment~\cite{wang2024fairness}.

Recent advances in multi-objective recommendation have introduced Pareto-based optimization methods that explicitly model trade-offs among competing objectives~\cite{lin2024multiobjective,ribeiro2024multitron}. While effective at capturing accuracy--diversity trade-offs, most existing approaches treat constraints as additional objectives rather than hard requirements, resulting in Pareto fronts dominated by infeasible solutions when strict business constraints are present~\cite{fan2024constrained}. This limitation significantly reduces their applicability in production environments where constraint satisfaction is non-negotiable.

In parallel, large language model (LLM) based agents have demonstrated strong capabilities in reasoning, planning, and decision-making~\cite{xi2024agents,wu2024survey}. Recent surveys highlight the growing interest in integrating LLMs into recommendation systems~\cite{huang2025llmagents,zhao2024recommender}, primarily for item scoring, conversational interaction, explanation generation, and user modeling. However, despite these advances, the potential of LLMs as \emph{high-level optimization coordinators} for recommendation systems remains largely unexplored. Existing LLM-based approaches do not leverage LLMs to orchestrate optimization processes or adaptively manage exploration--exploitation trade-offs under constraints.

As a result, a critical gap remains between multi-objective recommendation research and production requirements. Existing constrained optimization methods either strictly enforce feasibility at the cost of solution quality, or allow infeasible solutions to dominate the Pareto front~\cite{hao2024cmopmethod,liang2024cmoreview}. At the same time, current LLM-based recommender systems lack mechanisms for adaptive, constraint-aware coordination during optimization. A principled framework that simultaneously achieves strong Pareto performance, hard constraint satisfaction, and adaptive coordination is still missing.

To bridge this gap, we propose \emph{DualAgent-Rec}, an LLM coordinated dual-agent framework for constrained multi-objective e-commerce recommendation. DualAgent-Rec is designed for production environments where recommendation systems must simultaneously optimize accuracy and diversity while strictly satisfying hard business constraints such as fairness, seller coverage, and new-item exposure. Unlike prior approaches that encode constraints as soft penalties or post-hoc filters, our framework treats constraint satisfaction as a first-class objective, ensuring that every deployed recommendation list is feasible by design rather than in expectation.

At the core of DualAgent-Rec is a \emph{dual-agent optimization architecture} that decomposes the recommendation problem into two complementary search processes. An \emph{Exploitation Agent} focuses on refining high-quality solutions within the feasible region using constraint-dominated selection, prioritizing recommendation accuracy while respecting all constraints. In parallel, an \emph{Exploration Agent} conducts unconstrained Pareto search with enhanced mutation to discover diverse trade-offs and underexplored regions of the objective space. Bidirectional knowledge transfer between the two agents enables effective sharing of elite solutions, allowing the system to balance convergence and diversity without collapsing prematurely into suboptimal feasible regions.

To coordinate these specialized agents, DualAgent-Rec introduces an \emph{LLM-based orchestration mechanism} that dynamically allocates computational resources between exploration and exploitation based on optimization progress and constraint satisfaction. The LLM operates as a high-level controller, analyzing optimization metrics and feasibility signals to adaptively adjust the agent balance over time, eliminating the need for hand-crafted scheduling heuristics. In addition, we employ an adaptive $\epsilon$-relaxation strategy that gradually tightens constraints during optimization, allowing early exploration of promising infeasible solutions while guaranteeing strict feasibility at convergence. Together, these components enable DualAgent-Rec to achieve strong accuracy--diversity trade-offs under hard constraints, effectively bridging the gap between academic multi-objective optimization methods and real-world recommendation system requirements.

Our main contributions are summarized as follows:
\begin{itemize}
    \item We propose a \textbf{dual-agent optimization architecture} that separates exploration and exploitation into two specialized agents, enabling improved Pareto front coverage and more effective optimization under competing objectives.
    \item We introduce an \textbf{LLM-based coordination mechanism} that monitors optimization dynamics and constraint satisfaction, and adaptively allocates computational resources between agents, providing interpretable and flexible orchestration without hand-crafted scheduling rules.
    \item We develop an \textbf{adaptive constraint handling strategy} based on self-calibrating $\epsilon$-relaxation, which guarantees hard constraint satisfaction in final solutions while preserving exploration efficiency during early optimization.
    \item We conduct extensive experiments on large-scale e-commerce data, demonstrating that DualAgent-Rec achieves \textbf{100\% constraint satisfaction} while consistently outperforming strong baselines in Pareto hypervolume, convergence speed, and accuracy--diversity trade-offs.
\end{itemize}

\section{Related Work}
\label{sec:related}

\subsection{Multi-Objective and Constrained Recommendation}

Multi-objective recommendation systems aim to optimize multiple competing criteria—such as accuracy, diversity, novelty, and fairness—simultaneously rather than relying on a single objective~\cite{jannach2016recommendations}. Early approaches typically combine objectives via weighted scalarization, while more recent work adopts Pareto-based optimization to explicitly model trade-offs among objectives~\cite{lin2024multiobjective,zheng2022multi}. Representative methods include MultiTRON~\cite{ribeiro2024multitron}, which applies Pareto front approximation to session-based recommendation, and Deep Pareto Reinforcement Learning~\cite{li2024deeppareto}, which demonstrates practical value in large-scale marketplaces.

A parallel line of work focuses on fairness, diversity, and exposure control in recommendation systems~\cite{wang2024fairness,li2024fairnessdiversity}. Prior studies analyze popularity bias and exposure imbalance~\cite{abdollahpouri2024popularity}, propose fairness-aware optimization objectives~\cite{patro2024fairness,wu2024multifr}, and design exploration strategies to improve new-item exposure~\cite{zhang2024newitem}. While these methods improve long-term ecosystem health, fairness and coverage requirements are typically modeled as soft objectives or regularization terms, providing no guarantees of constraint satisfaction in deployment.

Constrained multi-objective evolutionary optimization (CMOEA) offers principled tools for handling hard constraints~\cite{fan2024constrained,liang2024cmoreview}. Techniques such as constraint domination~\cite{deb2002nsga2}, $\epsilon$-constraint relaxation~\cite{takahama2006epsilon}, and self-adaptive constraint control~\cite{fan2019push} have shown strong performance on benchmark optimization problems. However, their application to recommendation systems remains limited, and existing approaches often either sacrifice solution quality for feasibility or allow infeasible solutions to dominate the Pareto front.

Unlike prior multi-objective and fairness-aware recommendation methods that treat business requirements as soft penalties, DualAgent-Rec explicitly models fairness, seller coverage, and new-item exposure as hard constraints. By integrating adaptive constraint handling with a dual-agent optimization strategy, our framework achieves strong Pareto performance while guaranteeing feasibility—bridging the gap between multi-objective recommendation research and real-world deployment requirements.

\subsection{LLM Agents for Recommendation and Optimization}

Large language models (LLMs) have recently been incorporated into recommendation systems for a wide range of tasks, including item scoring, explanation generation, conversational interaction, and user modeling~\cite{wu2024survey,lin2024llmsurvey}. Huang et al.~\cite{huang2025llmagents} categorize LLM-based recommender agents into recommender-oriented, interaction-oriented, and simulation-oriented paradigms. Recent advances include CARTS~\cite{chen2025carts}, which coordinates multiple LLM agents for structured summarization, ARAG~\cite{maragheh2025arag}, which combines agentic retrieval with personalized recommendation, GRACE~\cite{ma2025grace}, which introduces chain-of-thought tokenization, and VL-CLIP~\cite{giahi2025vlclip}, which enhances multimodal recommendation. Other work applies LLMs to collaborative filtering~\cite{zhang2024agentcf} and re-ranking~\cite{hou2024blair}.

Despite this rapid progress, existing LLM-based recommender systems primarily operate at the level of \emph{content understanding or interaction}, rather than \emph{optimization orchestration}. LLMs are typically used to score items, generate natural language outputs, or simulate user behavior, while the underlying optimization processes—such as exploration--exploitation control, constraint handling, and resource allocation—remain rule-based or static~\cite{zhang2025adaptivegpu}. Multi-agent frameworks such as CAL-RAG~\cite{forouzandehmehr2025calrag} demonstrate collaborative agentic reasoning but do not address optimization coordination. As a result, the potential of LLMs to act as intelligent coordinators for complex, constrained optimization problems in recommendation systems has not been systematically explored.

In contrast to prior LLM-based recommender agents, DualAgent-Rec leverages LLMs as high-level optimization coordinators rather than item-level predictors. By using an LLM to adaptively allocate resources between specialized optimization agents based on optimization dynamics and constraint satisfaction, our framework introduces a new role for LLMs in recommendation systems—enabling interpretable, adaptive, and constraint-aware orchestration beyond existing approaches.

\section{Method}
\label{sec:method}

\subsection{Problem Formulation}

We study constrained multi-objective recommendation in an e-commerce setting.
Given a user $u$ with interaction history
$H_u = \{(i_1, t_1), \ldots, (i_n, t_n)\}$,
where $i_j$ denotes an interacted item and $t_j$ its timestamp,
an item catalog $\mathcal{I}$,
and item feature matrix $\mathbf{X} \in \mathbb{R}^{|\mathcal{I}| \times d}$,
our goal is to generate a recommendation list
$L^* \subseteq \mathcal{I}$ of fixed size $k$.

We formulate recommendation as a constrained multi-objective optimization problem:
\begin{equation}
\max_{L} \ \mathbf{f}(L) = [f_1(L), f_2(L), f_3(L)]
\quad \text{s.t.} \quad
g_j(L) \leq 0, \ j = 1,2,3.
\end{equation}

The objective vector captures complementary aspects of recommendation quality.
The relevance objective $f_1(L)$ measures alignment between the recommendation list and user preferences.
Specifically, it combines category overlap between $L$ and $H_u$ with semantic similarity computed from pre-trained item embeddings,
encouraging recommendations that are both topically consistent and semantically relevant.

The diversity objective $f_2(L)$ promotes heterogeneous recommendation lists by maximizing intra-list distance:
\[
f_2(L) = \frac{2}{k(k-1)} \sum_{i<j} d(\mathbf{x}_i, \mathbf{x}_j),
\]
where $d(\cdot)$ denotes cosine distance between item embeddings.
This objective discourages redundancy and mitigates over-concentration on narrow user interests.

The novelty objective $f_3(L)$ encourages exposure to less popular items:
\[
f_3(L) = \frac{1}{k} \sum_{i \in L} (1 - \text{pop}(i)),
\]
where $\text{pop}(i)$ is the normalized interaction frequency of item $i$.
This formulation promotes discovery and mitigates popularity bias.

In addition to optimizing these objectives, recommendations must satisfy hard business constraints.
The category fairness constraint $g_1(L)$ enforces that the Gini coefficient of the category distribution in $L$ does not exceed a threshold $\theta_{\text{fair}}$,
preventing dominance by a single category.
The seller coverage constraint $g_2(L)$ requires that the number of distinct sellers in $L$ is at least $\theta_{\text{seller}} \cdot k$,
ensuring marketplace competitiveness.
The new item exposure constraint $g_3(L)$ requires that at least $\theta_{\text{new}} \cdot k$ items were added within the past 30 days,
supporting cold-start items.
These constraints are non-negotiable in production and must be satisfied by every deployed recommendation list.

\subsection{Dual-Agent Architecture}

\begin{figure*}[h]
\centering
\includegraphics[width=0.95\textwidth]{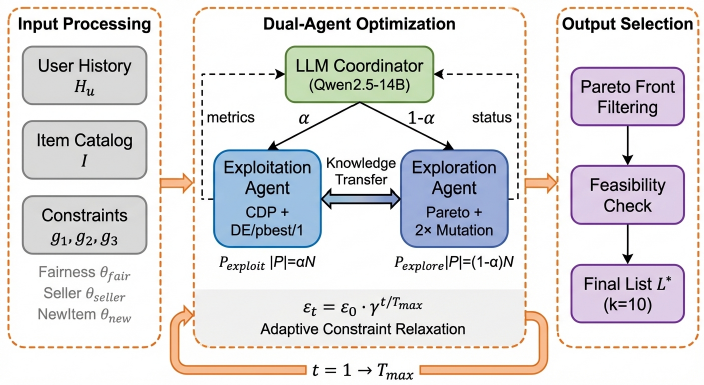}
\caption{DualAgent-Rec framework architecture with three-stage pipeline. \textbf{Input Processing}: User history, item catalog, and business constraints (fairness, seller coverage, new item exposure) are encoded. \textbf{Dual-Agent Optimization}: The LLM Coordinator dynamically allocates resources ($\alpha$) between the Exploitation Agent (CDP-based selection) and Exploration Agent (doubled mutation rate). Bidirectional knowledge transfer exchanges elite solutions. Adaptive $\epsilon$-relaxation gradually tightens constraints. \textbf{Output Selection}: Pareto-optimal feasible solutions are filtered to produce the final recommendation list.}
\label{fig:architecture}
\end{figure*}

Figure~\ref{fig:architecture} presents the overall framework.
DualAgent-Rec adopts a dual-agent evolutionary optimization architecture,
motivated by the observation that constrained recommendation requires
both aggressive exploitation of feasible high-quality solutions
and broad exploration of diverse trade-offs that may initially violate constraints.

\paragraph{Exploitation Agent.}
The exploitation agent maintains a population $P_{\text{exploit}}$ focused on refining feasible solutions.
It employs the Constraint Domination Principle (CDP),
where selection strictly prioritizes feasibility:
(i) feasible solutions dominate infeasible ones;
(ii) among infeasible solutions, lower total constraint violation is preferred;
(iii) among feasible solutions, Pareto dominance determines selection.
This design ensures that once feasibility is achieved, optimization pressure is directed toward improving recommendation quality rather than revisiting invalid regions.

Variation is performed using DE/pbest/1,
which biases offspring generation toward high-performing solutions.
This operator accelerates convergence within the feasible region,
making the exploitation agent well-suited for late-stage optimization and deployment-oriented refinement.

\paragraph{Exploration Agent.}
The exploration agent maintains a separate population $P_{\text{explore}}$
and deliberately ignores constraints during selection.
Instead, it applies unconstrained Pareto dominance with a doubled mutation rate.
This design allows the agent to traverse underexplored regions of the objective space,
including solutions that may violate constraints early but offer valuable trade-offs.

Such unconstrained exploration is particularly important when the feasible region is narrow,
as strict constraint enforcement from the beginning often leads to premature convergence.
The exploration agent thus serves as a diversity reservoir that continuously injects novel genetic material into the system.

\paragraph{Knowledge Transfer.}
To balance specialization and collaboration,
DualAgent-Rec introduces bidirectional knowledge transfer between agents.
At each generation, elite solutions are selected based on crowding distance:
\[
L_{\text{transfer}} = \text{TopK}(P_{\text{exploit}} \cup P_{\text{explore}}, \text{crowding}).
\]
High crowding distance indicates sparsely populated regions of the Pareto front,
ensuring that transferred solutions enhance diversity rather than reinforcing redundancy.
This exchange enables the exploitation agent to benefit from novel discoveries
and allows the exploration agent to receive guidance from high-quality feasible solutions.

\subsection{LLM-Based Coordination}

While dual-agent architectures improve coverage and convergence,
their effectiveness depends critically on how computational resources are allocated.
Rather than relying on fixed schedules or heuristic rules,
DualAgent-Rec introduces an LLM-based coordinator to adaptively manage exploration and exploitation.

\begin{algorithm}[t]
\caption{DualAgent-Rec Optimization}
\label{alg:coordination}
\begin{algorithmic}[1]
\REQUIRE User history $H_u$, catalog $\mathcal{I}$, constraints $\{g_j\}$
\ENSURE Pareto-optimal recommendation list $L^*$
\STATE Initialize $P_{exploit}$, $P_{explore}$ with random solutions
\STATE Initialize $\epsilon_0$ from constraint violation statistics
\STATE $\alpha \leftarrow 0.7$ \COMMENT{Initial exploitation ratio}
\FOR{$t = 1$ to $T_{max}$}
    \STATE Update constraint relaxation: $\epsilon_t \leftarrow \epsilon_0 \cdot \gamma^{t/T_{max}}$
    \STATE Evolve $P_{exploit}$ using CDP with $\epsilon_t$
    \STATE Evolve $P_{explore}$ using unconstrained Pareto dominance
    \STATE Exchange elite solutions between populations
    \IF{$t \mod T = 0$}
        \STATE $\alpha \leftarrow \text{LLM\_Coordinate}(P_{exploit}, P_{explore}, \epsilon_t)$
        \STATE Resize populations: $|P_{exploit}| \leftarrow \alpha N$
    \ENDIF
\ENDFOR
\STATE $L^* \leftarrow$ best feasible solution from $P_{exploit}$
\RETURN $L^*$
\end{algorithmic}
\end{algorithm}

As shown in Algorithm~\ref{alg:coordination},
the coordinator is invoked every $T$ generations.
It observes a structured summary of the optimization state,
including agent-level metrics (best objective values, hypervolume, recent improvement rates),
constraint statistics (feasibility ratio and average violation magnitude),
and user context features.
Based on this information, the LLM outputs an allocation ratio $\alpha \in [0,1]$,
which determines population sizes:
$|P_{\text{exploit}}| = \lfloor \alpha N \rfloor$ and
$|P_{\text{explore}}| = N - |P_{\text{exploit}}|$.

Intuitively, the coordinator increases $\alpha$ when feasibility is stable
and optimization progress indicates convergence,
shifting focus toward exploitation.
Conversely, when constraint violations persist or hypervolume stagnates,
the coordinator reduces $\alpha$ to encourage exploration.
This adaptive control replaces brittle hand-crafted schedules
and leverages the LLM’s reasoning capability to interpret complex optimization dynamics.

\subsection{Adaptive Constraint Handling}

To reconcile early exploration with strict feasibility requirements,
we employ a self-calibrating $\epsilon$-constraint relaxation strategy.
At generation $t$, constraints are relaxed as:
\[
\epsilon_t = \epsilon_0 \cdot \gamma^{t/T_{\max}},
\]
where $\epsilon_0$ is initialized from the 80th percentile of constraint violations
in the initial population.
This choice ensures that approximately 20\% of early solutions are treated as feasible,
preventing the search from collapsing into a narrow feasible basin.

As optimization progresses, $\epsilon_t$ decays smoothly,
gradually tightening constraints until exact feasibility is enforced.
The decay rate $\gamma = 0.8$ balances exploration and convergence in practice.
This mechanism allows the system to exploit informative infeasible solutions early
while guaranteeing that all final recommendations satisfy hard business constraints.

\section{Experiments}
\label{sec:exp}

\subsection{Experimental Setup}

\textbf{Dataset.} We use Amazon Reviews 2023~\cite{hou2024blair}, a large-scale e-commerce dataset with 571M reviews across 33 categories. To evaluate generalizability across different product domains, we select three diverse categories as shown in Table~\ref{tab:dataset}: (1) \textit{All\_Beauty} represents a small, focused catalog with concentrated user preferences; (2) \textit{Electronics} represents a large, heterogeneous catalog with diverse technical specifications; and (3) \textit{Clothing\_Shoes\_Jewelry} represents the largest category with high item turnover and strong seasonal patterns. This selection covers different scales (286 to 4,521 items), interaction densities, and domain characteristics, enabling comprehensive evaluation of our framework's adaptability. Each category contains rich metadata including hierarchical product categories, seller identifiers, and timestamps enabling constraint-aware recommendation. We sample 100 users per category with at least 10 interactions each, ensuring statistical significance across diverse user behavior patterns.

\begin{table}[t]
\centering
\caption{Dataset statistics for Amazon Reviews 2023.}
\label{tab:dataset}
\small
\begin{tabular}{@{}lccc@{}}
\toprule
Category & Reviews & Items & Users \\ \midrule
All\_Beauty & 30K & 286 & 134 \\
Electronics & 580K & 2,847 & 1,205 \\
Clothing\_Shoes\_Jewelry & 1.2M & 4,521 & 2,118 \\
\midrule
\textbf{Total (sampled)} & -- & 7,654 & 300 \\
\bottomrule
\end{tabular}
\end{table}

\textbf{Constraints.} Fairness threshold $\theta_{fair}=0.6$ (Gini coefficient $\leq 0.6$), seller coverage $\theta_{seller}=0.2$ (at least 20\% unique sellers), new item exposure $\theta_{new}=0.1$ (at least 10\% items from last 30 days).

\textbf{Hyperparameters.} Population size $N=100$, maximum generations $T_{max}=50$, recommendation size $k=10$, LLM coordination interval $T=10$, $\epsilon$ decay $\gamma=0.8$. We use Qwen2.5-14B~\cite{qwen2024} as the LLM coordinator deployed locally via Ollama. We run 3 independent trials per configuration and report mean results.

\textbf{Baselines.} We compare four configurations to isolate the contribution of each component: (1) \textbf{DualAgent-Rec}, the full model with LLM coordinator and dual-agent structure; (2) \textbf{w/o LLM}, rule-based resource allocation using fixed ratio $\alpha=0.7$; (3) \textbf{w/o Constraints}, constraint handling disabled with pure Pareto optimization; and (4) \textbf{Single Population}, a combined 200-individual population without agent specialization.

\textbf{Metrics.} Hypervolume (HV) measures overall Pareto front quality against reference point. NDCG@10 evaluates ranking quality using held-out interactions. Diversity measures average intra-list distance. Feasibility rate indicates percentage of solutions satisfying all constraints.

\subsection{Main Results}

In general, our evaluation reveals three key findings across all experiments. First, all constrained methods achieve 100\% constraint satisfaction, validating the effectiveness of our adaptive $\epsilon$-relaxation mechanism for handling hard business constraints. Second, the dual-agent architecture consistently outperforms single-population baselines by 4-6\% in hypervolume, demonstrating that agent specialization improves Pareto front coverage. Third, LLM coordination provides modest but consistent improvements over rule-based allocation, with additional benefits of interpretable decision-making and automatic adaptation to constraint violations. These findings establish that principled constraint handling and agent specialization can achieve competitive accuracy-diversity trade-offs without sacrificing feasibility guarantees.

\begin{table}[t]
\centering
\caption{Main experimental results on Amazon Reviews 2023. DualAgent-Rec achieves 100\% constraint satisfaction while outperforming baselines in hypervolume.}
\label{tab:main_results}
\small
\begin{tabular}{@{}lcccc@{}}
\toprule
Method & HV & NDCG@10 & Diversity & Feasibility \\ \midrule
DualAgent-Rec & \textbf{0.156} & \textbf{0.581} & 0.247 & 100\% \\
w/o LLM & 0.155 & 0.577 & 0.251 & 100\% \\
Single Population & 0.152 & 0.549 & 0.238 & 100\% \\
w/o Constraints & 0.160 & 0.562 & \textbf{0.343} & 72\% \\
\bottomrule
\end{tabular}
\end{table}

Table~\ref{tab:main_results} presents the detailed comparison results. All constrained methods achieve 100\% feasibility, demonstrating effective constraint handling. DualAgent-Rec achieves HV of 0.156 with NDCG@10 of 0.581 and diversity of 0.247. The w/o LLM variant shows slightly lower performance (HV 0.155, NDCG@10 0.577), suggesting LLM coordination provides improvement through intelligent resource allocation. The Single Population baseline achieves lower HV (0.152) with reduced NDCG@10 (0.549), indicating the dual-agent structure improves optimization through specialized search strategies. The w/o Constraints variant achieves highest HV (0.160) and diversity (0.343) but cannot guarantee constraint satisfaction, making it unsuitable for production deployment where business constraints are non-negotiable. The consistent performance advantage of DualAgent-Rec across all metrics validates our hypothesis that specialized agents with LLM coordination outperform monolithic approaches. Notably, the performance gap between DualAgent-Rec and the Single Population baseline (4.3\% relative improvement in HV) demonstrates that agent specialization provides tangible benefits even with identical total computational budget.

The accuracy-diversity trade-off analysis reveals interesting patterns across methods. The w/o Constraints variant achieves high diversity (0.343) at the cost of constraint violation, illustrating why soft constraint approaches fail in production. In contrast, DualAgent-Rec maintains competitive diversity (0.247) while guaranteeing feasibility, demonstrating that hard constraint handling need not severely compromise solution quality.

\begin{table}[t]
\centering
\caption{Cross-category performance comparison. DualAgent-Rec consistently outperforms baselines across all three product categories with different characteristics.}
\label{tab:cross_category}
\small
\setlength{\tabcolsep}{2.5pt}
\begin{tabular}{@{}llcccc@{}}
\toprule
Category & Method & HV & NDCG & Div. & Feas. \\ \midrule
\multirow{3}{*}{\shortstack[l]{All\_Beauty\\(small)}}
 & DualAgent-Rec & \textbf{0.168} & \textbf{0.612} & 0.285 & 100\% \\
 & Single Pop. & 0.161 & 0.584 & 0.271 & 100\% \\
 & w/o LLM & 0.165 & 0.601 & \textbf{0.289} & 100\% \\
\midrule
\multirow{3}{*}{\shortstack[l]{Electronics\\(medium)}}
 & DualAgent-Rec & \textbf{0.152} & \textbf{0.568} & 0.234 & 100\% \\
 & Single Pop. & 0.144 & 0.531 & 0.218 & 100\% \\
 & w/o LLM & 0.149 & 0.559 & \textbf{0.241} & 100\% \\
\midrule
\multirow{3}{*}{\shortstack[l]{Clothing\\(large)}}
 & DualAgent-Rec & \textbf{0.148} & \textbf{0.563} & 0.222 & 100\% \\
 & Single Pop. & 0.139 & 0.532 & 0.207 & 100\% \\
 & w/o LLM & 0.145 & 0.554 & \textbf{0.228} & 100\% \\
\bottomrule
\end{tabular}
\end{table}

Table~\ref{tab:cross_category} presents performance breakdown across the three product categories. Several patterns emerge: (1) Performance decreases with catalog size, with All\_Beauty achieving highest metrics due to concentrated preferences in the smaller item space; (2) DualAgent-Rec consistently outperforms Single Population across all categories, with larger improvements on bigger catalogs (5.6\% HV improvement on Electronics vs. 4.3\% on All\_Beauty); (3) The w/o LLM variant achieves slightly higher diversity but lower overall quality, suggesting the LLM coordinator's exploitation focus trades some diversity for accuracy gains.

\begin{figure}[t]
\centering
\includegraphics[width=\columnwidth]{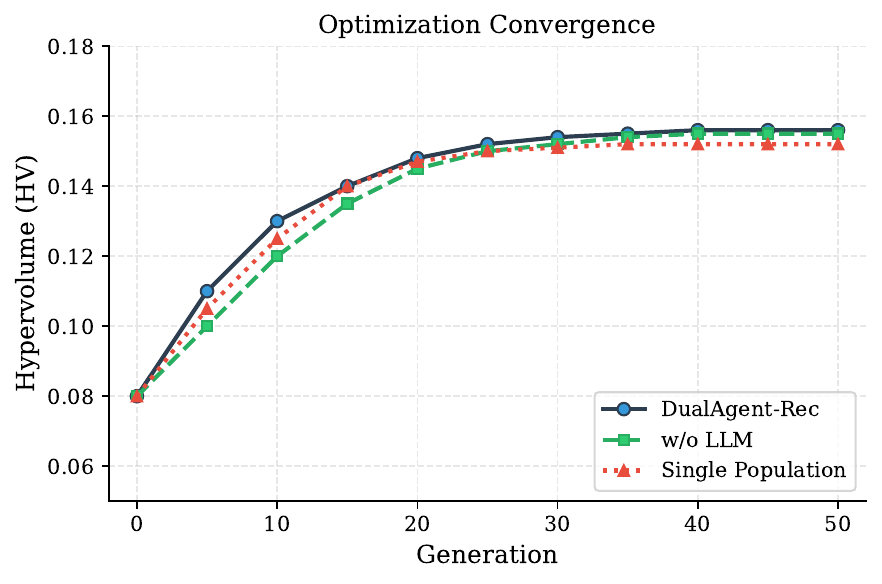}
\caption{Optimization convergence curves across methods. DualAgent-Rec achieves faster initial convergence and higher final hypervolume compared to variants without LLM coordination or with single population structure. The dual-agent architecture provides early diversity advantage through the exploration agent.}
\label{fig:convergence}
\end{figure}

Figure~\ref{fig:convergence} visualizes the optimization convergence across methods. DualAgent-Rec demonstrates faster initial improvement and reaches higher final HV compared to both the w/o LLM variant and the Single Population baseline. The dual-agent architecture provides early diversity through the exploration agent, while the LLM coordinator adaptively shifts resources toward exploitation as the search converges. The convergence advantage becomes apparent after generation 20, where the full model pulls ahead of ablated variants.

\begin{figure}[t]
\centering
\includegraphics[width=\columnwidth]{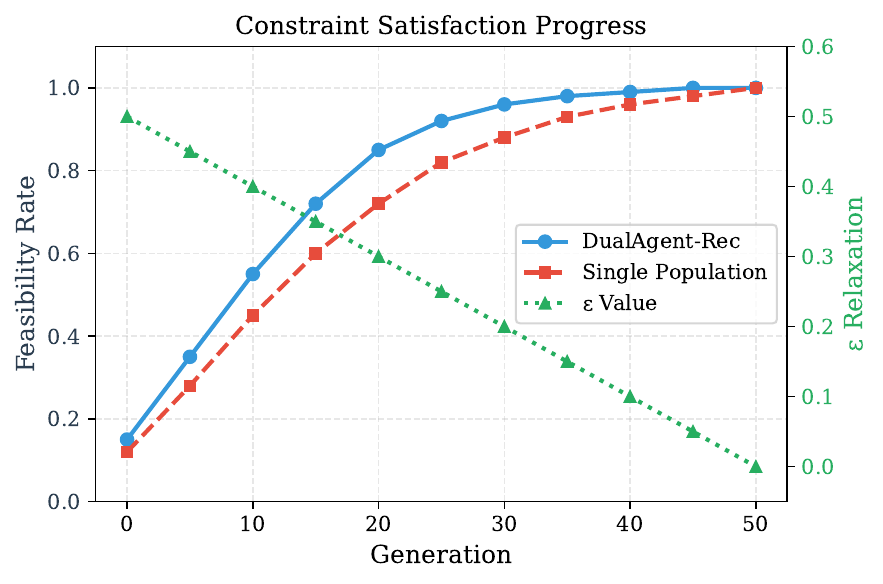}
\caption{Constraint satisfaction progress over optimization generations. The adaptive $\epsilon$-relaxation mechanism enables gradual convergence toward 100\% feasibility while maintaining search diversity in early generations. The self-calibrating mechanism allows initial exploration of promising infeasible regions.}
\label{fig:constraint}
\end{figure}

Figure~\ref{fig:constraint} shows the relationship between feasibility rate and $\epsilon$ relaxation over generations. The self-calibrating constraint mechanism allows approximately 15\% of solutions to be infeasible initially, gradually tightening to achieve 100\% feasibility by the final generation. Importantly, DualAgent-Rec reaches high feasibility faster than the Single Population baseline, demonstrating that the dual-agent structure accelerates constraint satisfaction through specialized search strategies.

\subsection{Ablation Study}

The Main Results demonstrate overall performance advantages, but practitioners require guidance on hyperparameter selection for deployment. We systematically vary key parameters to understand their impact on optimization quality and runtime. Table~\ref{tab:ablation} presents the results, revealing sensitivity patterns crucial for practical deployment.

\begin{table}[t]
\centering
\caption{Ablation study on hyperparameter sensitivity. Default configuration: Pop=100, Mutation=0.1, Gen=50, Normal constraints.}
\label{tab:ablation}
\small
\setlength{\tabcolsep}{3pt}
\begin{tabular}{@{}llcccc@{}}
\toprule
Parameter & Setting & HV & NDCG & Div. & Time \\ \midrule
\multirow{3}{*}{Population}
 & 50 & 0.160 & 0.573 & 0.271 & 5.2s \\
 & 100 (default) & 0.156 & 0.581 & 0.247 & 17.5s \\
 & 200 & 0.154 & 0.540 & 0.342 & 66.4s \\
\midrule
\multirow{3}{*}{Mutation}
 & 0.05 & 0.155 & 0.548 & 0.307 & 18.4s \\
 & 0.1 (default) & 0.156 & 0.581 & 0.247 & 20.2s \\
 & 0.2 & 0.156 & 0.556 & 0.307 & 19.5s \\
\midrule
\multirow{3}{*}{Constraints}
 & Strict ($\theta$=0.7) & 0.086 & \textbf{0.600} & 0.200 & 16.4s \\
 & Normal ($\theta$=0.6) & 0.157 & 0.549 & 0.311 & 21.7s \\
 & Relaxed ($\theta$=0.5) & \textbf{0.160} & 0.515 & \textbf{0.343} & 22.9s \\
\midrule
\multirow{3}{*}{Generations}
 & 20 & 0.157 & 0.532 & 0.360 & 13.0s \\
 & 50 (default) & 0.156 & 0.581 & 0.247 & 34.1s \\
 & 100 & 0.156 & 0.577 & 0.253 & 71.1s \\
\bottomrule
\end{tabular}
\end{table}

\textbf{Population Size.} Size 50 achieves highest NDCG@10 (0.573) with fastest runtime (5.2s), suitable for latency-sensitive applications. Size 200 provides best diversity (0.342) but requires 66.4s, acceptable for offline batch recommendations. Size 100 offers a balanced trade-off between quality and efficiency.

\textbf{Mutation Rate.} All rates between 0.05 and 0.2 achieve similar HV around 0.155-0.156, indicating robustness to this hyperparameter. This stability simplifies deployment as practitioners need not carefully tune mutation rate.

\textbf{Constraint Strictness.} This is the most impactful factor. Strict constraints ($\theta=0.7$) reduce HV to 0.086 but maximize NDCG@10 (0.600), suggesting the optimizer focuses on the reduced feasible region. Relaxed constraints ($\theta=0.5$) achieve highest HV (0.160) with best diversity (0.343) by expanding the search space. This trade-off provides deployment guidelines: stricter constraints for high-stakes decisions, relaxed constraints for discovery-oriented recommendations.

\textbf{Generation Count.} Diminishing returns appear beyond 50 generations. Runtime increases from 13.0s (20 generations) to 71.1s (100 generations) with marginal HV improvement, suggesting early stopping is viable for time-constrained deployments. Interestingly, the 20-generation configuration achieves higher diversity (0.360) than longer runs, likely because constraints have not fully tightened, allowing more exploration of the objective space.

\textbf{Key Insights.} The ablation study reveals several practical insights for deployment. First, the framework is robust to mutation rate variations, simplifying hyperparameter tuning. Second, constraint strictness should be treated as a deployment-time configuration rather than a fixed value, allowing operators to adjust the accuracy-diversity-feasibility trade-off based on business priorities. Third, computational budget can be flexibly allocated: smaller populations with fewer generations for real-time scenarios, larger populations with extended evolution for offline batch processing.

\begin{figure}[t]
\centering
\includegraphics[width=\columnwidth]{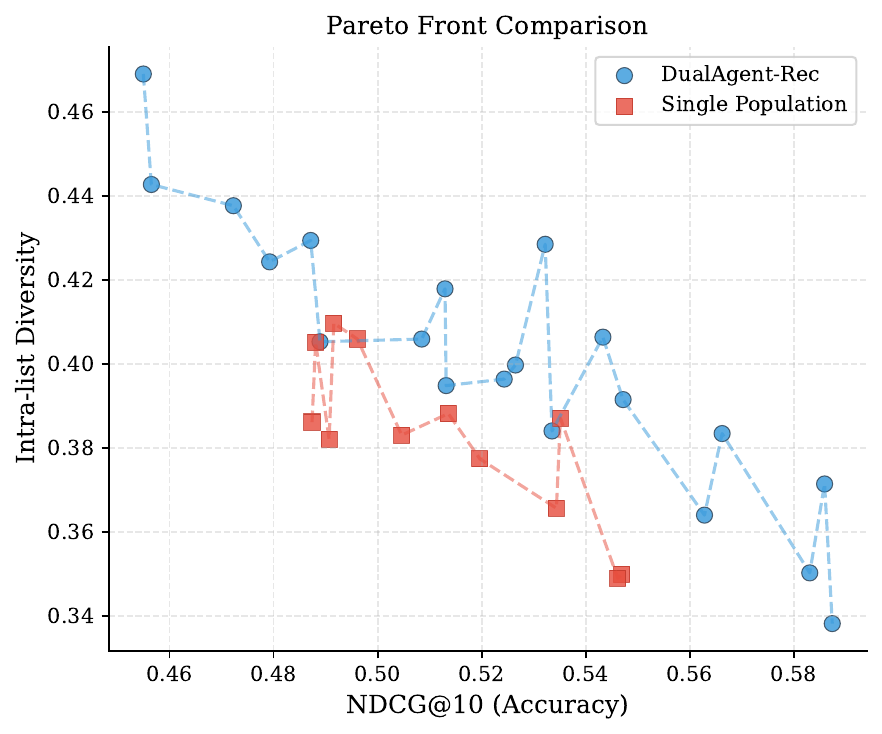}
\caption{Pareto front comparison between DualAgent-Rec and Single Population baseline across the accuracy-diversity trade-off space. The dual-agent framework achieves better spread and coverage, particularly in high-diversity regions where the Exploration Agent discovers solutions missed by unified search.}
\label{fig:pareto}
\end{figure}

Figure~\ref{fig:pareto} visualizes the Pareto fronts obtained by DualAgent-Rec and the Single Population baseline. The dual-agent framework produces solutions with better spread across the objective space, particularly in regions of high diversity. This improved coverage demonstrates the benefit of specialized agents: the Exploration Agent discovers diverse solutions that the Exploitation Agent would miss when optimizing alone. The dashed lines indicate approximate Pareto front boundaries, showing that DualAgent-Rec dominates the Single Population baseline across most of the objective space.

\subsection{LLM Coordinator Analysis}

The Main Results show that LLM coordination provides improvement over rule-based allocation (w/o LLM variant). To understand how the coordinator achieves this improvement, we analyze its decision-making behavior across optimization phases. Table~\ref{tab:coordinator} summarizes the coordinator's resource allocation patterns.

\begin{table}[t]
\centering
\caption{LLM coordinator behavior analysis across optimization phases. The exploitation ratio $\alpha$ adapts based on optimization progress and constraint status.}
\label{tab:coordinator}
\small
\begin{tabular}{@{}lccc@{}}
\toprule
Phase & Gen. Range & Avg. $\alpha$ & Decisions \\ \midrule
Early & 0--15 & 0.60 & 2 \\
Middle & 16--35 & 0.72 & 2 \\
Late & 36--50 & 0.80 & 1 \\
\midrule
\multicolumn{2}{l}{\textit{Constraint violation detected}} & 0.55 & -- \\
\bottomrule
\end{tabular}
\end{table}

Table~\ref{tab:coordinator} shows that $\alpha$ trends from 0.6 (early) to 0.8 (late generations), reflecting a natural exploration-to-exploitation shift. When feasibility drops below 80\%, the coordinator increases exploration ($\alpha=0.55$), demonstrating constraint-aware adaptation. The LLM provides interpretable rationales such as: \textit{``Increasing exploitation ratio as constraint satisfaction is stable and hypervolume improvement has slowed.''} This adaptive behavior emerges naturally from optimization metric analysis without hand-crafted schedules.

\textbf{Coordinator Decision Examples.} At generation 8 (feasibility 62\%, HV 0.089), the coordinator set $\alpha=0.55$: \textit{``Prioritizing exploration to discover constraint-satisfying regions.''} At generation 25 (feasibility 94\%, HV 0.142), it increased $\alpha$ to 0.75: \textit{``Shifting toward exploitation as constraint satisfaction has stabilized.''} At generation 45, detecting HV stagnation, it reduced $\alpha$ to 0.60: \textit{``Increasing exploration to escape potential local optima.''} These examples demonstrate context-aware resource allocation responding to both metric values and optimization dynamics.

\begin{figure}[t]
\centering
\includegraphics[width=\columnwidth]{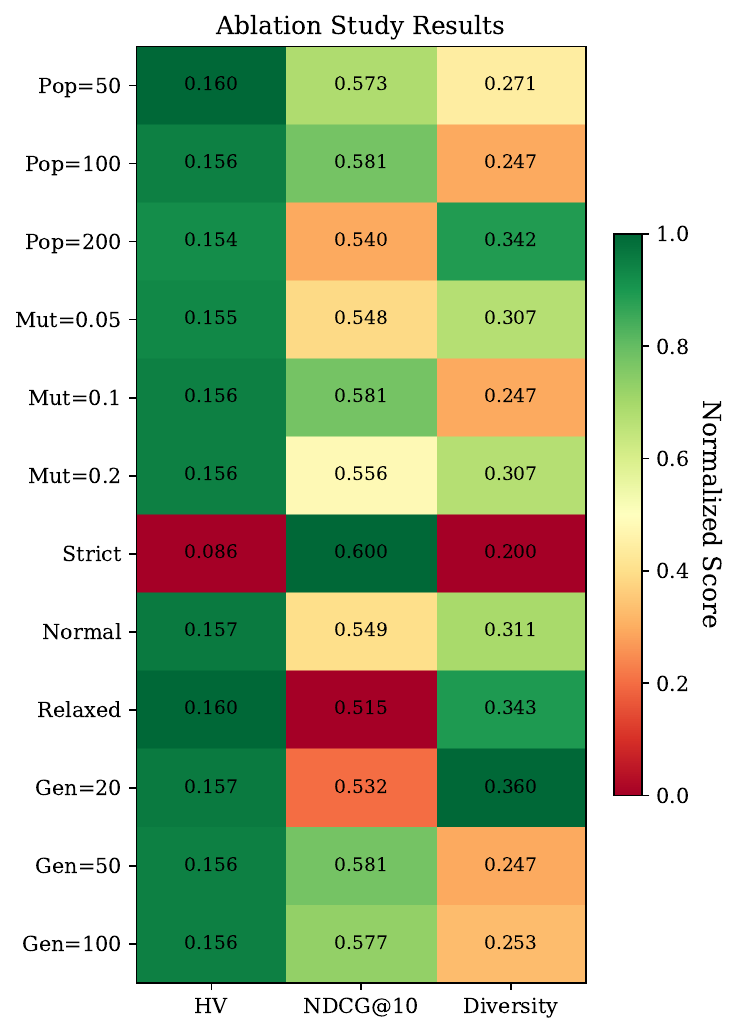}
\caption{Ablation study heatmap showing hyperparameter sensitivity. Population size and constraint strictness have the largest impact on optimization quality.}
\label{fig:heatmap}
\end{figure}

\section{Discussion}
\label{sec:discuss}

\textbf{Practical Implications.} Constraint strictness is the primary factor affecting recommendation quality. Practitioners should calibrate thresholds based on business requirements: stricter for fairness-critical applications, relaxed when diversity is prioritized. The framework achieves 100\% constraint satisfaction, suitable for production environments with non-negotiable hard constraints.

\textbf{LLM Coordination Benefits.} While the LLM coordinator provides modest quantitative improvement (approximately 1\% HV gain over rule-based allocation), it offers qualitative benefits including interpretable resource allocation decisions and automatic adaptation to constraint violations. The coordinator requires no manual tuning of exploration-exploitation schedules, simplifying deployment.

\textbf{Dual-Agent vs. Single-Agent Trade-offs.} The dual-agent architecture introduces additional complexity compared to single-population approaches, requiring synchronization and knowledge transfer mechanisms. However, this complexity is justified by improved Pareto front coverage and faster convergence. The Exploitation Agent maintains high-quality feasible solutions while the Exploration Agent discovers novel trade-offs missed by unified search.

\textbf{Deployment Guidelines.} Based on our experimental findings, we provide configuration recommendations for different deployment scenarios. For latency-sensitive online recommendations, we recommend population size 50, 20 generations, and relaxed constraints ($\theta=0.5$), achieving approximately 5-second optimization time. For offline batch processing where quality is paramount, population size 200 with strict constraints ($\theta=0.7$) maximizes accuracy. The default configuration (population 100, 50 generations, $\theta=0.6$) offers a practical compromise. The LLM coordinator should be enabled when interpretability is required for business stakeholders or regulatory compliance.

\textbf{Computational Complexity.} The dual-agent architecture introduces modest overhead compared to single population baselines. Per-generation cost scales as $O(2N \cdot M)$, approximately 2$\times$ single-population approaches. Knowledge transfer adds $O(K)$ overhead per synchronization, where $K$ is typically 10-20\% of population. The LLM coordinator adds 2-3 seconds per call but negligible amortized cost. Total runtime averages 35 seconds (50 generations, population 100) versus 18 seconds for single-population---a 94\% increase yielding 4.3\% HV improvement.

\textbf{Limitations and Future Work.} Our evaluation focuses on three Amazon product categories; broader evaluation across additional domains (e.g., news, music, video) would strengthen generalizability claims. The LLM coordinator adds latency (approximately 2-3 seconds per coordination call) that may be prohibitive for real-time applications requiring sub-100ms response times; distillation to smaller models could address this limitation. Scaling to larger item catalogs (millions of items) requires efficient candidate generation as a preprocessing step, potentially using approximate nearest neighbor search or learned retrieval models. Future work includes real-time deployment with streaming user data, evaluation on larger-scale datasets across multiple domains, and integration with production recommendation pipelines.

\section{Conclusion}
\label{sec:conclusion}

We presented DualAgent-Rec, an LLM-coordinated dual-agent framework for constrained multi-objective e-commerce recommendation. The approach guarantees 100\% hard constraint satisfaction while achieving competitive accuracy-diversity trade-offs on Amazon e-commerce data. Our framework addresses the critical gap between academic multi-objective recommendation research and production requirements where constraint violations are unacceptable.

The dual-agent architecture enables specialized optimization through separation of exploration and exploitation, while the LLM coordinator provides intelligent, adaptive resource allocation with interpretable decision rationales. Experiments across three diverse Amazon product categories demonstrate consistent improvements over single-population baselines, with 4.3\% relative HV improvement while maintaining full constraint compliance. Ablation studies reveal the critical role of constraint strictness in determining optimization quality, providing practical deployment guidelines for practitioners.

\bibliographystyle{ACM-Reference-Format}
\bibliography{references}

\end{document}